\documentclass[prl,preprint,showpacs,aps,amsmath]{revtex4}
\usepackage{graphicx}
\usepackage{color}
\usepackage{subfigure}

\begin{document}

\title{Breakdown of Landau Theory in Overdoped Cuprates near the Onset of Superconductivity}
\author{M. Ossadnik$^{1}$, C. Honerkamp$^{2}$, T.M. Rice$^1$, and M. Sigrist$^1$}
\affiliation{$^{1}$ Theoretical Physics, ETH Z\"urich, CH-8093 Z\"urich, Switzerland, \\
$^{2}$ Theoretical Physics, $U$niversit\"at W\"urzburg, D-97074 W\"urzburg, Germany}
\date{\today}

\pacs{71.10.Hf, 71.27.+a, 74.72.-h}

\begin{abstract}
We use the functional renormalization group to analyze the temperature dependence of the quasi-particle scattering rates in the two-dimensional Hubbard model below half-filling. Using a band structure appropriate to overdoped Tl$_2$Ba$_2$CuO$_{6+x}$  we find a strongly angle dependent term  linearly dependent on temperature which derives from an increasing scattering vertex as the energy scale is lowered. This behavior agrees with recent experiments and confirms earlier conclusions on the origin of the breakdown of the Landau Fermi liquid near the onset of superconductivity.
\end{abstract}
\maketitle

 Recently Abdel-Jawad and collaborators \cite{abdel} reported a striking correlation between charge transport and superconductivity in heavily overdoped high-temperature superconducting cuprates. They found that the onset of superconductivity with doping coincided with the appearance in the normal phase of strong anisotropic quasiparticle scattering. The application of a magnetic field to suppress the superconductivity revealed that the anisotropic term in the inplane transport scattering rate was linear in temperature violating the perturbative quadratic dependence characteristic of a Landau Fermi liquid. Earlier investigations of a 2-dimensional Hubbard model on a square lattice using a functional renormalization group (RG) method found that $d$-wave pairing in the overdoped region of the phase diagram was driven by the appearance at low energies and temperatures of a strongly anisotropic scattering vertex in the particle-particle and particle-hole channels \cite{zanchi,halboth,breakdown}. Further investigations revealed that the self-energy is also anisotropic\cite{zanchi2,katanin,ehdoping,rohe}. In this letter we present an extensive RG study of the doping and temperature dependence of the quasi-particle scattering rate with the pairing instability suppressed, and compare our results  with the experiments of Ref. \cite{abdel}. We find a rising anisotropic scattering vertex that gives an anisotropic contribution to the decay rate with a linear, not quadratic, temperature dependence when the pairing divergence is suppressed. We wish to stress that this breakdown of standard Landau Fermi liquid behavior is not associated with a divergent density of states from a van Hove singularity at the Fermi energy but instead is due to strong scattering processes at large momentum transfer which appear in the RG flows as a precursor to the Mott insulating behavior at half filling. The origin of anomalous decay rate in our calculations is quite different from that proposed by Metzner and co-workers\cite{dellanna} who put forward  a model based on  small angle scattering near to a Pomeranchuk instability.

The experiments by Abdel-Jawad and collaborators \cite{abdel} were carried on well characterized Tl$_2$Ba$_2$CuO$_{6+x}$ samples. Their studies of the interlayer angle-dependent magnetoresistance (ADMR) provided also detailed Fermi surface information which we use to fix the band parameters in the RG calculations. We use a moderate starting value of the onsite repulsion, $U$, in the one loop RG equations. Our goal is a qualitative rather than a quantitative description which would require a larger value of $U$ and multi-loop corrections to the RG flow equations. In the experiments a magnetic field is applied to suppress superconductivity and study the normal phase down to low temperatures. However it is technically very difficult to introduce a magnetic field into the RG calculations. Instead we introduce an elastic scattering term to suppress the $d$-wave pairing instability in the calculation of the RG flows of the scattering vertex. This vertex is then used as input into a standard lowest order calculation of the quasiparticle decay rate.

The kinetic energy applicable to Tl$_2$Ba$_2$CuO$_{6+x}$ takes the form 
\begin{eqnarray}
\epsilon(k_x,k_y) &=&-2t_1\left(\cos k_x + \cos k_y\right) +4t_2\left(\cos k_x \cos k_y\right)\nonumber\\
 && + 2t_3\left(\cos 2k_x + \cos 2 k_y\right) +4t_4(\cos 2k_x\cos k_y 
\nonumber\\ &&+ \cos 2k_y\cos k_x) + 4t_5\left(\cos 2k_x\cos 2k_y\right),
\end{eqnarray}
with $t_1=0.181,\;t_2=0.075,\;t_3=0.004,\;t_4=-0.010,\text{ and } t_5=0.0013 (\text{eV})$. \\

\begin{figure}
 \centering
\includegraphics[width=5cm]{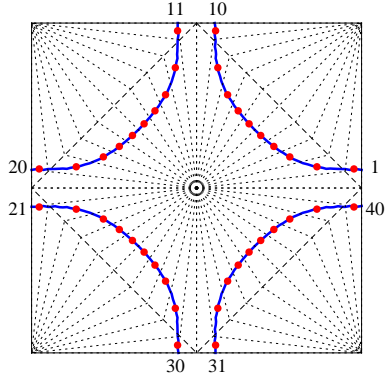}
\caption{(Color online) Fermi surface (solid line) and discretization of the BZ for $p=0.22$. The boundaries of the patches (labelled by $1,2,\ldots,40$) are indicated by the dashed lines. All vertices are evaluated at the points marked by the red dots, and are taken to be constant within each patch.}
\label{fig:patches}
\end{figure}

Our approach relies on the functional RG equation for the one-particle irreducible (1PI) generating functional $\Gamma[\Phi]$, which is derived in \cite{tnt,breakdown}, and which leads to a hierarchy of coupled flow equations for the 1PI vertices after a suitable expansion of the functional. We use a Wilsonian flow scheme with a sharp momentum cutoff. It was shown later that this method underestimates effects of small wavelength scattering \cite{tflow}, but also that these processes are important only if the Fermi surface is close to a van Hove singularity, which is not the case in the doping regime studied here. In order to solve the flow equations, the hierarchy has to be truncated, and in the following we will use the standard truncation of neglecting all vertices with more than four legs. In this approximation, the only quantities appearing in the calculation are the self-energy $\Sigma_\Lambda(k)$ and the 4-point vertices $V_\Lambda(k_1,k_2, k_3)$. The $k_i$ also contain the frequency, $k_i = \left(\omega_i,\mathbf{k}_i\right)$. All propagators contain a sharp infrared cutoff $\chi_\Lambda(\mathbf{k})=\theta(|\epsilon(\mathbf{k})|-\Lambda)$ in momentum space, where the flow parameter $\Lambda$ flows from $\Lambda=\infty$ to $\Lambda=0$ with the initial condition $V_\infty(k_1,k_2,k_3)=U$. As these equations are still too complicated to be solved, we introduce some further approximations by neglecting the frequency dependence of all vertices and by discretizing their momentum dependence. The latter is done by dividing the Brillouin zone into elongated patches each of which contains a part of the Fermi surface. The momentum dependence of the vertices is approximated by a step function which is constant in each patch. The vertices are calculated at a reference point in each patch, which we choose to lie where the Fermi surface crosses the center of the patch, as shown in FIG. \ref{fig:patches} for the case of a hole concentration $p=0.30$.\\

Typically, at low energies the truncated flows diverge at some finite energy scale $\Lambda$. The leading divergence can be interpreted as the dominant instability, and the scale at which the divergence occurs gives an estimate of the corresponding $T_c$ \cite{breakdown}. In the regime of interest here, $d$-wave superconductivity is the leading instability, and in our approximation $T_c$ takes the values $T_c = 0.26 t_1$ for $p=0.15$, $T_c=0.22t_1$ for $p=0.22$, and $T_c=0.16t_1$ for $p=0.30$. These temperatures are way too high, mainly because we neglect self-energy corrections in the flow of the scattering vertex. Nevertheless, $T_c$ grows with decreasing hole doping reproducing qualitatively the experimental results \cite{abdel}.

\begin{figure}
 \centering
\includegraphics[width=9cm]{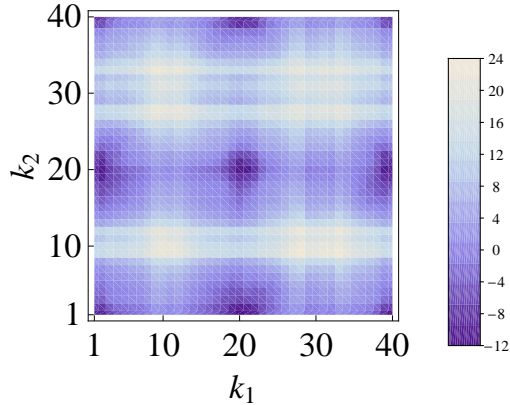} 
\caption{(Color online) Characteristic momentum dependence of the renormalized vertex $V_\Lambda(\mathbf{k}_1,\mathbf{k}_2,\mathbf{k}_3)/t_1$ for $p=0.22$, $T=0.02 t_1$, $1/\tau_0=0.2t_1$ at $\Lambda=0$. In the figure, the dependence on the two ingoing wave vectors $(\mathbf{k}_1, \mathbf{k}_2)$ is shown, where the outgoing wave vector $\mathbf{k}_3$ is taken to lie in patch $1$ close to $(\pi,0)$ (cf. Fig. \ref{fig:patches}) and $\mathbf{k}_4$ is fixed by momentum conservation.}
\label{fig:vanis}
\end{figure}

The experiments were carried out in a high magnetic field, which suppresses superconductivity and allows to access the normal state down to low temperatures. However, including a magnetic field into our RG calculation is very difficult, so that we choose a different way to suppress superconductivity, namely we introduce an isotropic scattering rate $1/\tau_0$ into the free part of the action. This smears out the Fermi distribution at the Fermi surface, which in turn regularizes the loop integrals and subsequently the flow of the 4-point vertices. This scattering rate will only be included in the flow equation for the 4-point vertex, whereas the flow equation for the self-energy is left unaltered. We found that for our choice of $U=4t_1$ and the range of temperatures ($T\geq 0.004 t_1$) and dopings ($p \geq 0.15$), a scattering rate of $1/\tau_0 = 0.8 t_1$ is sufficient to suppress the divergences associated with superconductivity, so that on average the vertices are comparable to the bandwidth and the largest vertices do not grow larger than $\approx3\times$bandwidth (Fig. \ref{fig:vanis}).

In FIG. \ref{fig:vanis} a typical result of our calculations for the 4-point vertex $V_\Lambda(\mathbf{k}_1,\mathbf{k}_2,\mathbf{k}_3)$ at energy scale $\Lambda$ is displayed for a fixed outgoing wavevector $\mathbf{k}_3$ close to $(\pi,0)$ as a function of the two incoming wavevectors $(\mathbf{k}_1,\mathbf{k}_2)$. The remaining outgoing wavevector is determined by momentum conservation allowing for umklapp processes. The strongest scattering processes occur for a momentum change of $(\pi,\pi)$. These scattering vertices are approximately $\propto \Lambda^{-1/2}$ for small values of the cutoff $\Lambda\sim T$. However, for very small cutoff, the square root divergence is suppressed due to the presence of the isotropic scattering rate $1/\tau_0$.

As we are only interested in the scattering rates at the Fermi surface, which are given by $\Im\Sigma(\mathbf{k}\in FS, \omega\rightarrow 0+i\delta)$, we will restrict the calculation of the self-energy to this quantity in the following. Obviously, the frequency-dependence of $\Sigma$ cannot be neglected in the calculation. On the other hand, if we neglect the frequency-dependence of the 4-point vertices, it is clear from the structure of the flow equations that $\Sigma$ will also be frequency-independent, as only Hartree and Fock diagrams are included. However, as shown earlier by one of us\cite{ehdoping}, one can overcome this difficulty by replacing the 4-point vertex appearing in the self-energy flow equation by the integrated flow equation of the vertex, schematically,
\begin{equation}
\begin{split}
 \Sigma_{\Lambda=0}&=\int d\Lambda V_\Lambda S_\Lambda  \\ &=\int d\Lambda\left(\int d\bar{\Lambda} V_{\bar{\Lambda}} S_{\bar{\Lambda}} G_{\bar{\Lambda}} V_{\bar{\Lambda}}\right) S_\Lambda,
\label{eq:twoloop}
\end{split}
\end{equation}
where in our approximation the single-scale propagator $S_\Lambda$ \cite{breakdown,tnt} and the full propagator $G_\Lambda$ are related to the free propagator $G_0$ by
\begin{equation}
 S_\Lambda = \dot{\chi}_\Lambda G_0 \text{ and } G_\Lambda = \chi_\Lambda G_0,
\end{equation}
respectively. The RHS of eq. (\ref{eq:twoloop}) depends on $\Lambda$ only through the cutoff $\chi_\Lambda$. After a partial integration with respect to $\Lambda$ and after explicitly inserting a sharp cutoff $\chi_\Lambda(\mathbf{k}) = \Theta(|\epsilon(\mathbf{k})|-\Lambda)$ we have
\begin{eqnarray}
 \Sigma_{\Lambda=0}&=& \int d\Lambda \theta(|\epsilon(\mathbf{k}_1)|-\Lambda)\delta(|\epsilon(\mathbf{k}_2)|-\Lambda)\theta(\Lambda-|\epsilon(\mathbf{k}_3)|) \nonumber\\ & & \times V_\Lambda^2 G_0(k_1)G_0(k_2)G_0(k_3),
\label{eq:sigmafinal}
\end{eqnarray}
and summation and integration over internal momenta and Matsubara frequencies is implied. Thus the first propagator has support above, the second at, and the third below the cutoff $\Lambda$. \\
\begin{figure}
\includegraphics[width=5cm]{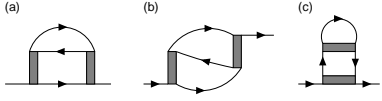}
 \caption{Two-loop diagrams contributing to the self-energy. Only diagrams a) and b) contribute to the scattering rates.}
\label{fig:sigmadiagrams}
\end{figure}
The diagrams corresponding to this equation for the self-energy are shown in FIG. \ref{fig:sigmadiagrams}. As we are interested in the scattering rates at the Fermi surface, we need only consider diagrams a) and b), because the contribution of diagram c) is real for external frequencies $\omega+i\delta$. For a) and b), for external frequency $\lim_{\omega\rightarrow 0}\omega+i\delta$, we obtain an imaginary part $\propto \delta\left(\epsilon(\mathbf{k}_3)-\epsilon(\mathbf{k}_2)-\epsilon(\mathbf{k}_1)\right)$, reflecting energy conservation.

It turns out that neglecting the flow of the 4-point vertices, i.e. setting $V_\Lambda=U$ in eq. (\ref{eq:twoloop}), is equivalent to a second order perturbative calculation of the scattering rate, which gives a $T^2$ behavior away from van Hove singularities. All deviations from the Landau theory scaling form may be attributed to the renormalization of the 4-point vertices.

\begin{figure}
\includegraphics[width=6cm]{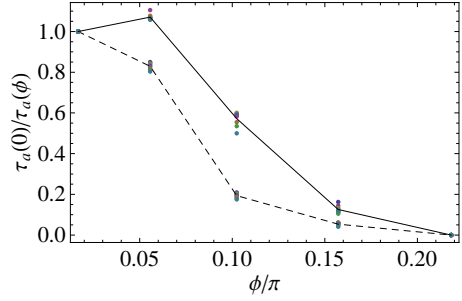}
\caption{(Color online) Angular dependence of the anisotropic component of the quasi-particle scattering rate on segment of the Fermi surface for $p=0.15$ (dashed line) and $p=0.30$ (solid line). The scattering rates are normalized to unity in the anti-nodal direction. The dots are the values at different temperatures, the lines are the temperature averaged angular dependence.}
\label{fig:angulardependence}
\end{figure}

Based on eq.\,(\ref{eq:sigmafinal}) we calculate both the temperature and the doping dependence of the angle-resolved quasi-particle scattering rates at the Fermi surface using the method explained above. We find that the scattering rates are anisotropic for all choices of parameters. The precise shape of the angle dependence changes with doping, but does not change very much with temperature, as shown in FIG. \ref{fig:angulardependence}. 
In general, we find that in the nodal direction ($\phi=\pi/4$) the scattering rates have a minimum, and increase towards the anti-nodal direction ($\phi=0$). The size of the anisotropy grows as doping is decreased. Calculations without the background scattering rates as a regulator showed that also $T_c$ increases with lower doping. This is in accord with the results of the magnetoresistance measurements carried out by Abdel-Jawad et al.\cite{abdel}, where it was found that with decreasing doping both $T_c$ and the anisotropic part of the scattering rates increase, whereas the uniform component remains constant.

Separating the scattering rates into an isotropic and an anisotropic part, we write
\begin{equation}
 \frac{1}{\tau}(\phi,T)= \frac{1}{\tau_i}(T) + \frac{1}{\tau_a}(\phi,T),
\end{equation}
where $1/\tau_i \equiv \min_\phi 1/\tau(\Phi)$ so that $1/\tau_a(\phi) \geq 0$. We characterize the $T$-dependence of the anisotropic part by averaging over the angle, 
\begin{equation}
\langle 1/\tau_a \rangle(T) = \frac{1}{2\pi} \int_0^{2\pi} d\phi \frac{1}{\tau_a}(\phi,T),
\end{equation}
which makes sense as the angular dependence of the anisotropic part is approximately independent of temperature (FIG. \ref{fig:angulardependence}).
\begin{figure}
\subfigure[]{\includegraphics[width=6cm]{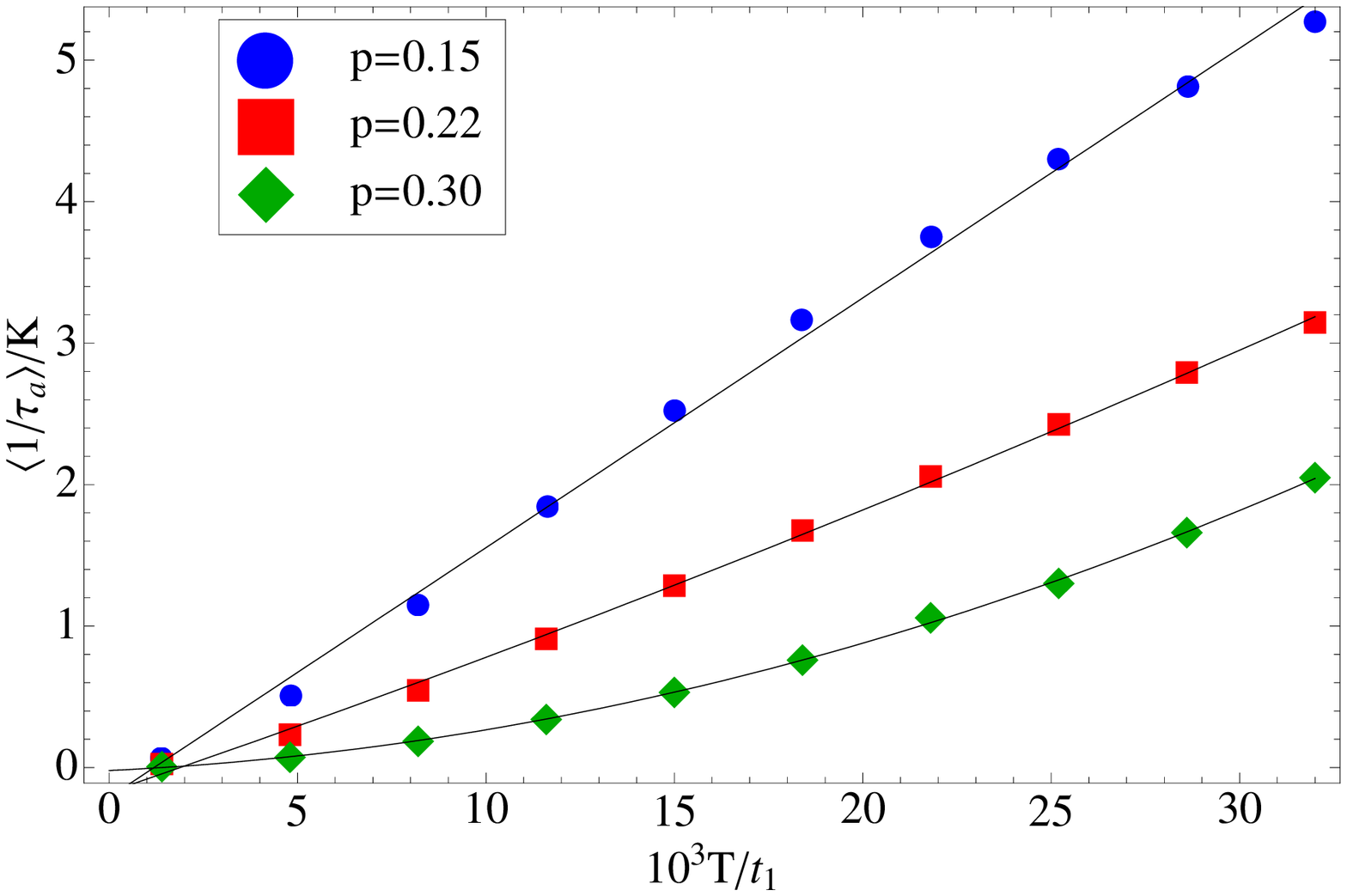}}
\subfigure[]{\includegraphics[width=6cm]{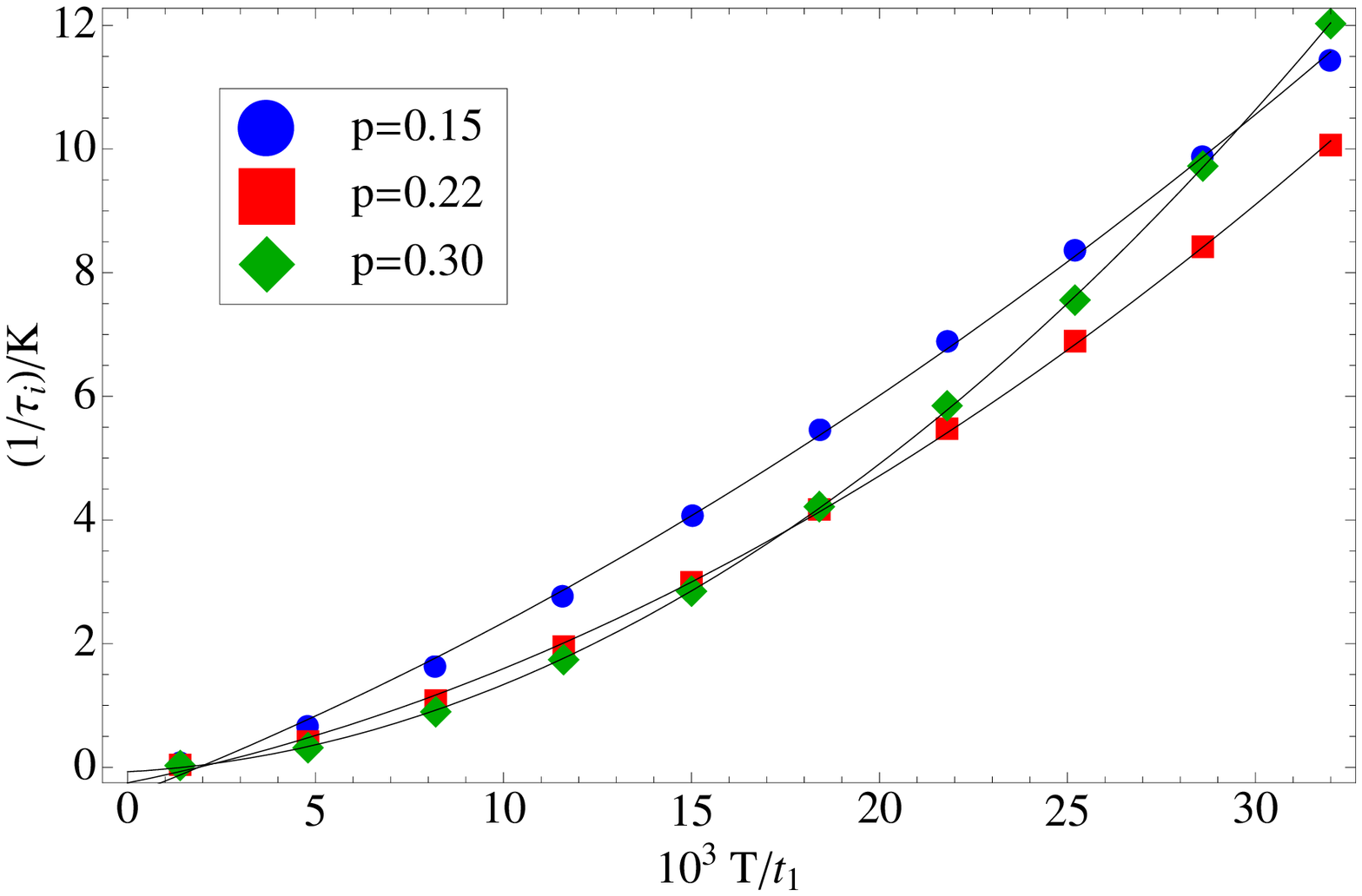}}
\caption{(Color online) Temperature dependence of (a) the anisotropic and (b) the isotropic component of the quasi particle scattering rate at the Fermi surface for different values of the hole doping. The solid lines are fits to quadratic polynomials in $T$.}
\label{fig:Tdependence}
\end{figure}
Using these definitions, we find that the $T$-dependence of $1/\tau_i$ and $\langle 1/\tau_a\rangle$ can be fitted very well by a quadratic polynomial, as shown in FIG. \ref{fig:Tdependence}. In the same figure, one sees that for $p \leq 0.25$, $\langle 1/\tau_a\rangle$ becomes \emph{linear} in $T$ with a coefficient which increases with decreasing hole doping, whereas the isotropic part $1/\tau_i$ is always dominated by a quadratic term which does not change much with doping. Thus our calculations reproduce the main features of the striking correlation between charge transport and superconductivity reported by Abdel-Jawad and coworkers.

Abdel-Jawad et al. \cite{abdel} also commented on the contrast between the transport properties of their overdoped samples which show the shortest lifetimes in the antinodal directions and the low temperature ARPES data showing shortest lifetimes in the nodal directions in the superconducting phase of overdoped Tl$_2$Ba$_2$CuO$_{6+x}$ samples\cite{plate}. However, Wakabayashi et al. \cite{wakabayashi} have attributed this latter behavior to a reduction in the elastic scattering rate near the antinodes due to a coherence effect which acts strongest at the (superconducting) gap energy. Multiple scattering in this case leads to a depression of the elastic scattering rate from impurities, defects etc. Such  processes are quite distinct to the intrinsic inelastic scattering processes in the normal state discussed here.

These results signal a clear breakdown in Landau-Fermi liquid behavior which leads to a universal $T^2$ dependence. We also wish to stress that this result is not due to a proximity to the van Hove singularity at the saddlepoint of the band structure. This lies well below the Fermi energy at energies $\gg T$. Further, the increase in the linear term in $T$ with decreasing hole dopings occurs as the energy of the van Hove singularity moves further away from the Fermi energy. In our calculations the breakdown of Landau theory arises from the increase in the 4-point vertex with decreasing energy scale. Note that this increase is not restricted to the $d$-wave pairing Cooper channel since the divergence in this channel is suppressed in our calculations. Examination of the RG flows shows that several channels in the 4-point vertex grow simultaneously, e.g. particle-hole and particle-particle umklapp processes. This phenomenon is not simply a precursor of $d$-wave superconductivity but rather signals that a crossover to strong coupling in several channels of the 4-point vertex is responsible for the breakdown of the Landau-Fermi liquid behavior. This simultaneous enhancement of several channels through mutual reinforcement was earlier identified as a key feature of the anomalous Fermi liquid in the cuprates and associated with the onset of resonant valence bond (RVB) behavior \cite{breakdown,laeuchli}. For the parameters used above, this crossover takes place at $p \approx 0.14$. However, due to the rich momentum dependence of the 4-point vertex, the nature of the strong coupling phase is unclear and requires further investigations.

In conclusion, the RG calculations presented here demonstrate that the anomalous behavior of the inplane quasi-particle scattering rate revealed by the ADMR experiments \cite{abdel} on overdoped cuprates can be found as an intrinsic feature of the doped Hubbard model already at weaker interaction strengths. Likewise, the positive correlation of the critical temperature for $d$-wave superconductivity with the strength of the scattering rate anisotropy comes out of the RG treatment without additional assumptions.

We are grateful to N. Hussey, L. Taillefer, A. Katanin, and W. Hanke for useful discussions. This study was financially supported by the Swiss National fonds through the NCCR MaNEP. CH acknowledges financial support by the DFG research unit 538.

\end{document}